# Physical energy cost serves as the "invisible hand" governing economic valuation: direct evidence from the U.S. metal market

Zhicen Liu[1,2], Joel Koerwer[3], Jiro Nemoto[4] & Hidefumi Imura[5]

*[1]Department of Biological Science, [2]Centre for gene research, [3]Department of Physics, [4]Graduate School of Economics and [5]Graduate School of Environmental Studies, Nagoya University, Furo-cho, Chikusa-ku, Nagoya 464-8602, Japan.*

**Corresponding author:**

Zhicen Liu
Phone: +81-52-789-4527; Fax: +81-52-789-4526;
Address: Center for Gene Research, Nagoya University, Furo, Chikusa, Nagoya 464-8602, Japan;
E-mail: zhicen@gene.nagoya-u.ac.jp

**Abstract**

Energy supply is mandatory for the production of economic value. Nevertheless, tradition dictates that an enigmatic "invisible hand" governs economic valuation. Physical scientists have long proposed alternative but testable energy cost theories of economic valuation, and have shown the gross correlation between energy consumption and economic output at the national level through input-output energy analysis. However, due to the difficulty of precise energy analysis and highly complicated real markets, no decisive evidence directly linking energy costs to the selling prices of individual commodities has yet been found. Over the past century, the US metal market has accumulated a huge body of price data, which for the first time ever provides us the opportunity to quantitatively examine the direct energy-value correlation. Here, by analyzing the market price data of 65 purified chemical elements (mainly metals) relative to the total energy consumption for refining them from naturally occurring geochemical conditions, we found a clear correlation between the energy cost and their market prices. The underlying physics we proposed has compatibility with conventional economic concepts such as the ratio between supply & demand or scarcity's role in economic valuation. It demonstrates how energy cost serves as the "invisible hand" governing economic valuation. Thorough understanding of this energy connection between the human economic and the Earth's biogeochemical dimensions is essential for improving the overall energy efficiency and furthermore the sustainability of the human society.

**1. Introduction**

Inflation, unemployment, stock prices, and economic growth, in fact all aspects of the human economy deteriorate when energy prices surge. Not only have wars been waged over energy resources, but their use lies at the root of anthropogenic climate change. Understanding the energy basis underlying the human economy may be more critical for improving the overall energy efficiency and sustainability of humanity than any single technological innovation or political decision.

Physical scientists have long noticed the indispensable role energy plays in economic systems (Soddy, 1933; Georgescu-Roegen, 1971; Costanza 1980, 2004; Odum, 1996; Ayres, 1998). In 1886, Boltzmann suggested that life is primarily a struggle over available energy (Costanza 1980). Soddy later studied the flow of energy

......bodydoneunderlying the economy and proposed the energy cost theory for economic valuation in 1933(Soddy, 1933; Costanza 1980). Previous input-output energy analyses have revealed an aggregate correlation between energy consumption (embodied energy) and economic output at the national level (Costanza 1980, 2004; Cleveland et al., 1984). Nevertheless, little is known about the direct link between the energy cost of production and the selling prices of individual commodities. Because the correlation between energy cost for production and the selling prices of individual commodities on the real market is extremely complicated, Costanaza admitted (2004), "Given, on the one hand, the enormous data requirements to calculate energy costs accurately and, on the other hand, the pervasive market imperfections complicating market prices, there is no unambiguous correct answer."

Over the past century, the US metal market has accumulated a huge body of price data. Such abundant data for the first time provided us the opportunity to quantitatively examine the direct energy-value correlation of individual commodities: the purified elements (mainly metals). The following sections report our analysis on the market price data of 65 purified chemical elements (from 1959 to 1998) relative to the total energy consumption for refining them from naturally occurring geochemical conditions. Based on the unambiguous energy-value correlation shown by the analysis, we further discuss how physical energy cost underlies conventional economic concepts, such as the ratio between supply & demand or scarcity's role in economic valuation.

## 2. Material and Methods

Purified chemical elements are ideal subjects for investigating the direct energy-value correlation. Physically, their energy states can be precisely defined from

4physical and geochemical data (Emsley, 1998; Reimann and Caritat, 1998; Rudnick and Gao, 2003). They are the simplest commodities, consisting of only a single component.

Economically, their market values are well documented. Many elements (iron, copper etc.), having been traded on industrial scales for more than a century, provide a large dataset for detailed statistical analysis. The 65 individual elements we investigated also provide an adequately large sample to verify the generality of the energy-value correlation. Moreover, for most of human history, purified metals served as the foundation of economic value systems. The units for economic value, be it Dollar, Pound or Yen, all originally represented the amount of elemental gold, silver, etc. (Table.1). Today, these metals are still major indicators of monetary value; therefore, the elements' prices bear directly upon the energy-value correlation.

We collected the price data from the U.S. Bureau of Mines and the U.S. Geological Survey on all chemical elements having trading records on the US metal market (Fig.1) (Saches 1999; U.S. Geological survey 1996, 1999, 2003, 2005; Bureau of Mines 1961, 1993). Annual prices are the averages for the year whenever available to avoid short-term fluctuations. To make the prices comparable through history, the constant 1992 dollar was set as the unit of market value (Saches 1999).

The prices for barium, boron, manganese, potassium, strontium and sodium, which were unavailable from the U.S. Bureau of Mines and the U.S. Geological Survey data are from other references (Earnshaw and Greenwood 1997; Emsley 1998) and were treated as market prices for 1998. Historical prices of some of the elements were those of slightly different purities (Saches 1999). The standard free-energy changes during extraction were adapted from absolute electronegativities (Emsley 1998).

## 3. Results

When refined from their naturally occurring geological state, the elements' economic values change from nearly zero to their market prices. Simultaneously, their physical energy states also change while being concentrated during the refining processes (Fig.2). The free energy change of the elements due to the concentrating process, $\Delta G$, is given by the equation $\Delta G = RT \ln \frac{[A]_{pure}}{[A]_{env}}$, where $[A]_{pure}$ is the mole concentration of the traded refined form of the element under room conditions ($\approx$295 K, 1 atm) (Earnshaw and Greenwood 1997; Emsley 1998; Lide 2004), $[A]_{env}$ is the mole concentration of the element in the natural environment which is the geochemical abundance in the upper continental crust (Earnshaw and Greenwood 1997; Emsley 1998; Lide 2004; Rudnick and Gao 2003; Reimann and Caritat 2003), $R$ is the gas constant, and $T$ is the absolute temperature. The fact that highly valued precious metals undergo greater energy changes upon being refined from their very low geological abundances $[A]_{env}$, prompted us to investigate how energy changes relate to economic valuation of the elements. In addition, the concentration change $\frac{[A]_{pure}}{[A]_{env}}$, physically defines the concept of "scarcity", which has hitherto only been connected intuitively with economic valuation.

Figure 3 shows the correlation (coefficient$\approx$0.53) between the free energy change $\Delta G$ and the logarithm of the selling price. This correlation, as if kept steady by an "invisible hand", holds over the 40-year period covered in this study and the entire range of the elements' prices (over seven orders from $10^{-3}$ to $10^4$ 1992US$ per mole).

Apart from being concentrated, the elements were also released from their natural binding chemicals during refining, changing their energy state by $\Delta G°$ (standard free-energy change, see methods). By assigning colours to the elements according to their level of standard free energy change $\Delta G°$, we see further confirmation of the energy-value correlation. In Fig.3, the elements are clearly stratified based on their $\Delta G°$: the alkali and alkaline earth elements, which have the greatest $\Delta G°$, possess much greater prices than the other elements having similar "scarcity"(concentration change $\frac{[A]_{pure}}{[A]_{env}}$). They occupy the upper left region of the plot. The elements having intermediate values of $\Delta G°$ are in the middle of the plot alongside the best-fit line of the energy-value correlation. The elements with low $\Delta G°$ occupy the lower right half of the plot.

The varied refining processes for different elements have only two purposes: one is to overcome $\Delta G$, that is to concentrate the elements by mining the ores from the bedrock, fragmenting, and transporting them, etc.; the other is to overcome $\Delta G°$, that is to release the elements from unwanted binding chemicals by using such as reductive reactants or electrolysis. Therefore, the total energy cost $\sum E$ for refining the elements can be expressed as

$$\sum E = a\Delta G + b\Delta G° \qquad (1)$$

where $a\Delta G$ is the energy cost for concentrating, and $b\Delta G°$ is the energy cost for releasing chemical bindings. Energy analysis of the copper industry has shown that energy efficiency for the concentrating processes is approximately one order lower than that for releasing the chemical bindings (Ayres et al., 2003). In Fig.4 we show, by selecting the efficiency ratio $a/b=13$, a high correlation (correlation coefficient>0.80) between total energy costs $\sum E$ and the selling prices of the elements. The energy-value correlation is maintained throughout the investigated period (1959-1998), despite the differing refining processes between elements and over time. This is expected by the nature of the energy-value correlation: if the selling price of an element is much higher than the general energy-value correlation, it has the potential to bring more than average profit for each unit of energy invested. This provides incentive for technological innovation and the entry of competitors, by which the price is eventually brought down to the general energy-value level.

The physics underlying the logarithmic energy-value or precisely energy-price (value/unit) correlation is straightforward, because monetary value originated from and still closely relates to the amount of elemental gold or silver etc. Price is the amount of money paid for exchanging per unit volume of a given commodity, therefore $Price = \frac{Money}{Volume} \approx [Au]$, according to equation (1) the energy cost for refining the element gold at a given concentration is $\sum E = aRT \ln \frac{[Au]}{[Au]_{env}} \approx aRT \ln \frac{Price}{[Au]_{env}}$, where $b\Delta G°$ has been omitted, because it is negligibly small for gold. Therefore, $\sum E$ is proportional to the logarithm of price.

## 4. Discussion

The energy-value correlation integrates well with the classical concepts of scarcity and the ratio of supply and demand as the basis of economic valuation. The ratio of concentrations between refined and naturally occurring forms, $\frac{[A]_{pure}}{[A]_{env}}$, embodies the concept of scarcity. If demand depletes the local supply, $[A]_{env}$, the price will increase due to greater scarcity $\frac{[A]_{pure}}{[A]_{env}}$. The earth's upper crust, acting as an enormous reservoir in the biogeochemical cycle of the elements, stabilizes the price over the long term. For this reason, the energy-value correlation holds over long periods, as shown in Fig.4.

The physics of economic valuation sets the baseline about which human influence causes price fluctuation. While, normally, economics focuses on price fluctuations that span much less than one order of magnitude and elapse within months, our study deals the more fundamental question of why in the first place the prices of the different elements vary over a range of seven orders of magnitude but remain at their relative levels over decades. For complex commodities, this energy-value correlation may be distorted beyond recognition by the multi-step stochastic fluctuations of real markets. Fortunately, previous research has circumvented this difficulty by indirectly showing the aggregate correlation between domestic energy consumption and the total economic value of produced commodities (Gross Domestic Product) (e.g., Costanza 1980; Cleveland et al., 1984). Combining this result with ours builds a strong case for the energy cost origin of economic valuation long proposed by physical scientists (Soddy, 1933; Georgescu-Roegen, 1971; Costanza 1980, 2004; Cleveland et al., 1984; Odum, 1996; Ayres, 1998).

Like in the old proverb, in which a single elephant is taken one aspect at a time, natural scientists and economists sometimes form very different or even opposite opinions on economic and environmental issues. The picture we suggest of a more unified approach to the biogeochemical and economic dimensions will provide a common platform for researchers from different disciplines to cooperate in overcoming humanity's formidable sustainability challenges.

Facing the uncertainty of future energy supplies and environmental consequences of fossil fuels, quantifying the energy basis of the economy is urgently needed to systematically improve its energy efficiency. Further studies are essential for rational decision making in everything from international issues to personal life style and consumer choice under ever-stringent energy and carbon budgets.

**Acknowledgements**

We thank M. Ishiura, S. Ito and K. Chan for critical reading and discussion about the manuscript; K. Ishii and S. Kondo for suggestions about the manuscript; O. Sakura for encouragement at the initiation of this research. We also thank the anonymous referees for their kind advice. This work is supported by the Ministry of Education, Culture, Sports, Science and Technology, Japan, and the Ito Foundation.

U.S. Geological Survey. *Minerals Yearbook, Vol. 1, Metals and Minerals, 2002* (United States Government Printing Office, Washington, D.C., 2003).

**Figure 1. The investigated elements and their energy-value correlation**
Prices are from data of 1998, unit is in 1992 constant US dollar.

**Figure 2. The energy-value correlation** After being purified from their naturally occurring states, both the energy state and price of an element change. We studied the correlation between these two changes.

**Figure 3. Correlation between market value and the Gibbs energy change for each mole of the elements** The market prices of the elements are correlated (correlation coefficient=0.53) to their change of Gibbs free energy during the processes of refinement. The elements are coloured according to their standard free energy change $\Delta G°$. Prices are from data of 1998, unit is 1992 US dollar.

**Figure 4. Correlation between total energy cost for refinement and the market value of each mole of the elements** total energy cost for refinement $\sum E$ closely correlate to the market value of the elements (correlation coefficient>0.80)When a/b=13 in equation (1). The correlation is held through the investigated 40-year period. Prices are in 1992 constant US dollar.

Table 1 **Metal based value system (European)**

| Ratio | Roman Empire | France | UK | Italy | Notes |
|---|---|---|---|---|---|
| 240 | Libra(327.48g) | Livre | Pound | Lira | Libra = £ = Pound |
| 12 | Solidus(16.37g) | Sol | Shilling | Soldo | Solidus = $ = Dollar |
| 1 | Denarius(1.36g) | Denier | Penny | Denaro | 1~1.5g ( 8-15th century. ) |

Zhicen_Liu_ fig.1

Zhicen_Liu_fig2

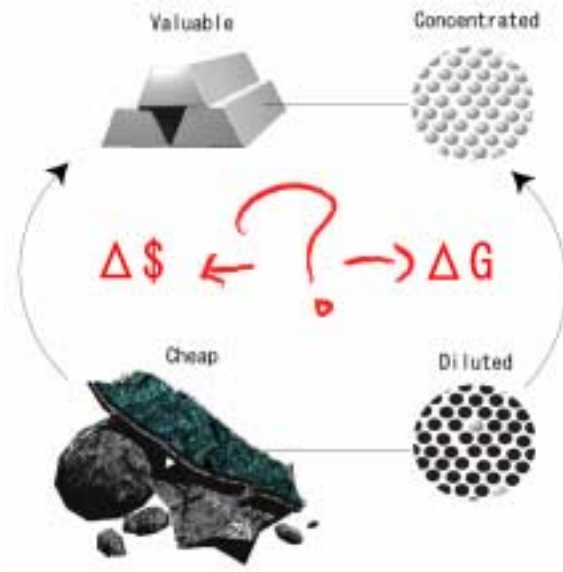

**Figure-3**
**Click here to download Figure: fig-3.doc**

Zhicen_Liu_fi3

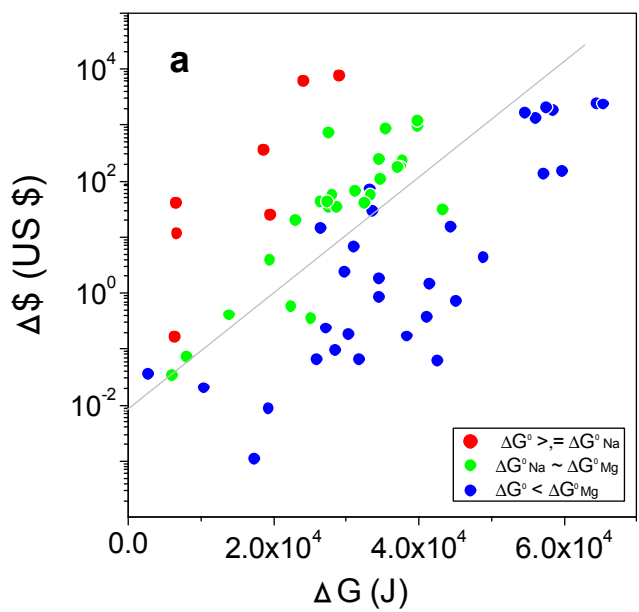

Zhicen_Liu_fig4

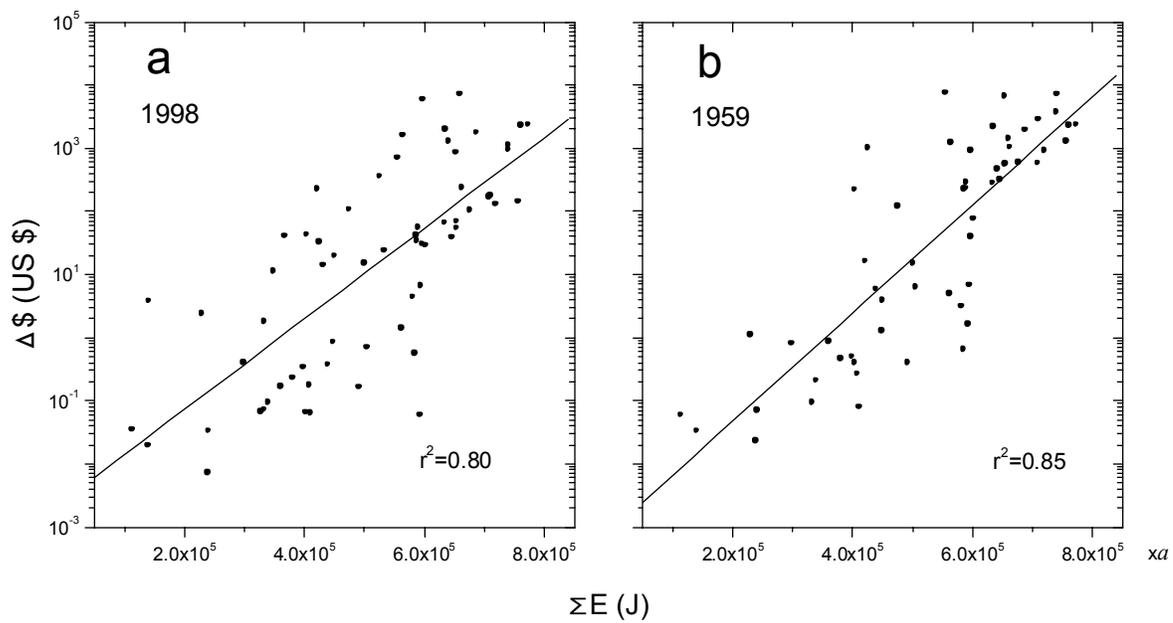